\documentclass[nofootinbib,prd]{revtex4}%
\usepackage{graphicx}
\usepackage{mathrsfs}
\usepackage{yfonts}
\usepackage{tipa}
\usepackage{bm}
\usepackage{bbm}
\usepackage{amsmath}
\usepackage{amssymb}
\usepackage{amsthm}
\usepackage{hyperref}
\usepackage{amsfonts}
\usepackage{mathtools}
\usepackage{latexsym}
\usepackage[greek,english]{babel}
\usepackage{booktabs}
\usepackage{xcolor}
\usepackage[iso-8859-7]{inputenc}%
\begin{document}
\title{Short-range approximation to Casimir wormholes inspired by scalar and electric fields}
\author{Remo Garattini}
\email{remo.garattini@unibg.it}
\affiliation{Universit\`a degli Studi di Bergamo, Dipartimento di Ingegneria e Scienze
Applicate, Viale Marconi 5, 24044 Dalmine (Bergamo), Italy  and I.N.F.N.-sezione di Milano, Milan, Italy.}
\author{Athanasios G. Tzikas}
\email{athanasios.tzikas@unibg.it}
\affiliation{Universit\`a degli Studi di Bergamo, Dipartimento di Ingegneria e Scienze
Applicate, Viale Marconi 5, 24044 Dalmine (Bergamo), Italy.}

\begin{abstract}

We investigate a static traversable wormhole sustained by a combination of a minimally coupled scalar field and an electric field, with exotic matter sourced by Casimir energy. Considering two scenarios, where the Casimir plate separation is either radially variable or fixed, we derive   analytical near-throat solutions for both massless and  massive scalar fields. To ensure consistency of the field equations, a thermal tensor is also incorporated, consisting solely of pressure terms that vanish at the throat. In all cases, we obtain well-behaved wormhole manifolds with throat sizes that scale proportionally with the number of elementary charges the wormhole can support.

\end{abstract}
\maketitle

\section{Introduction}
\label{intro}

It is well established that a Traversable Wormhole (TW) requires exotic matter to be sustained. Since exotic matter has yet to be discovered, one may seek an alternative that behaves in a similar manner. In this context, the Casimir effect presents a promising candidate for mimicking exotic matter \cite{Casimir}. The reason is the following: if two plane, parallel, closely spaced, uncharged metallic plates are placed in a vacuum, an attractive force arises between them. This force is generated by a negative energy, represented by
\begin{equation}
E^{\text{Ren}}(  d )  =-\frac{\hbar c\pi^{2}S}{720d^{3}} \,
\end{equation}
where $S$ is the surface of the plates, $d$ is the distance between them and
$E^{\text{Ren}}(  d )  $ stands for the renormalized energy. This energy arises from the zero-point energy  of the quantum electrodynamic field. A noteworthy property is that it suggests the presence of an equation of state of the form
$P=\omega\rho$ with $\omega=3\,$, where $P$ stands for the pressure and $\rho$ for the
energy density. Based on this property, a specific type of TW, termed as the Casimir Wormhole \cite{CW}, has been investigated. Its geometric profile is described by the following functions
\begin{equation}
    \Phi(r)   ={\ln}\left(  {\frac{3r}{3r+r_{0}}}\right)  \qquad \mathrm{and} \qquad b(r)   =\frac{2}{3}r_{0}+\frac{r_{0}^{2}}{3r} \label{CW}
\end{equation}
with $\Phi\!\left(r\right)  $ and $b(r)$ representing the redshift
 and the shape function of the wormhole respectively. Such a wormhole solution satisfies the Einstein Field Equations (EFE), provided that an additional inhomogeneous equation of state is introduced between the transverse pressure $p_{t}(
r)  $ and the energy density $\rho (  r )$, namely
\begin{equation}
p_{t}(r)=\omega_{t}\left(  r\right)  \rho\left(  r\right)  .%
\end{equation}
Casimir wormholes have attracted considerable attention in recent years. Generalizations of them have been explored  by introducing temperature effects into the original Casimir result \cite{HCW,Cha24}, replacing the parallel-plate configuration with cylindrical plates \cite{MeZ24}, combining them with a scalar field \cite{RGAZ} or a static electric field \cite{STC22,CCW}, and incorporating rotations \cite{RCW0,RCW}. Solutions have also been investigated within the frameworks of modified theories of gravity, such as the Brans-Dicke theory \cite{BD}, $F(R,\mathcal{L}_{m})-$gravity\cite{FRL,FRL2}, and Einstein Gauss-Bonnet gravity \cite{EGB,EGB2}. Further developments include Yukawa deformations \cite{YC,YC2,YC3} and Generalized Absurdly Benign Traversable Wormholes \cite{GABTW}\footnote{For a review on this subject, see also Ref.~\cite{MG17} and references therein.}.
Nevertheless, a combination of the three physical sources, namely the Casimir effect, electric field, and scalar field, has not been studied before.  For this reason, we undertake such an investigation here. Let us also point out that our analysis focuses on a scalar field with the appropriate polarity, in contrast to well-established frameworks \cite{Kim,BV,BV1,Butcher} that rely on scalar fields with the opposite polarity$-$commonly referred to as \textit{phantom fields}$-$which can also act as sources for traversable wormholes even in the absence of an additional Casimir contribution.  Such configurations lead to the well-known Ellis-Bronnikov  TW solution \cite{Ellis, Bronnikov}. 

The purpose of this paper is to investigate the conditions under which these three physical sources can coexist. To proceed, we introduce the following spacetime metric in natural units
\begin{equation}
\mathrm{d}s^{2}=-e^{2\Phi\left(  r\right)  }\,\mathrm{d}t^{2}+\frac{\mathrm{d}r^{2}}{1-b(r)/r}%
+r^{2}\,(\mathrm{d}\theta^{2}+\sin^{2}{\theta}\,\mathrm{d}\varphi^{2})\, \label{metric}%
\end{equation}
representing a spherically symmetric and static wormhole \cite{MT,Visser}. Here, $\Phi\!\left(
r\right)  $ and $b(r)$ are arbitrary functions of the radial coordinate
$r\in\left[  r_{0},+\infty\right)  $. The minimum value $r=r_0$ represents the wormhole throat.  An important
property of a TW is the flare-out condition, given by
$(b(r)-r b^{\prime}(r))/b^{2}(r)>0\,$. The prime symbol ($'$) denotes from now on differentiation with respect to the radial coordinate $r$. This must be satisfied along with the condition
$1-b(r)/r>0\,$. Furthermore, at the throat, the condition $b(r_{0})=r_{0}$ must hold, and
the additional requirement $b^{\prime}(r_{0})<1$ is  imposed to obtain well-behaved wormhole solutions.
It is imperative that  no horizons are present, which are
identified as the surfaces with $e^{2\Phi\left(  r\right)  }\rightarrow 0\,$. As a result,
 $\Phi\left(  r\right)  $ must be finite everywhere. 
 
Now with the help of
the line element  \eqref{metric}, we can write down the following set representing the EFE 
\begin{equation}
\frac{b^{\prime}\left(  r\right)  }{r^{2}}=\kappa\rho\left(  r\right)  
\label{rho}%
\end{equation}%
\begin{equation}
\frac{2}{r}\left(  1-\frac{b\left(  r\right)  }{r}\right)  \Phi^{\prime
}(  r)  -\frac{b\left(  r\right)  }{r^{3}}=\kappa p_{r}\left(
r\right)   \label{pr}%
\end{equation}%
\begin{equation}
\left(  1-\frac{b\left(  r\right)  }{r}\right)  \left[  \Phi^{\prime\prime
}\!\left(  r\right)  +\Phi^{\prime}(  r )  \left(  \Phi^{\prime
}(  r)  +\frac{1}{r}\right)  \right]  -\frac{b^{\prime}(
r )  r-b\left(  r\right)  }{2r^{2}}\left(  \Phi^{\prime}(
r)  +\frac{1}{r}\right)  =\kappa p_{t}(r) \label{pt}%
\end{equation}
in which $\kappa=8\pi G\,$, $\rho (  r )  $ is the total energy density, $p_{r}(
r)  $ is the total  radial pressure, and $p_{t} (  r )  $ is the total
tangential pressure. We can complete the EFE with the expression of the
conservation of the Stress-Energy Tensor (SET) which can be written in the same
 reference frame as
\begin{equation} \label{SET_cons}
p_{r}^{\prime}\left(  r\right)  =\frac{2}{r}\left(  p_{t}\left(  r\right)
-p_{r}\left(  r\right)  \right)  -\left(  \rho\left(  r\right)  +p_{r}\left(
r\right)  \right)  \Phi^{\prime}\left(  r\right)  \,.
\end{equation}
 The rest of the paper is organized as follows: In Sec.~\ref{sec2}, we analyze the structure of the SET of the sources under investigation. The total SET of our system consists of four parts: one resulting from the Casimir setup, one from the scalar field, one from the electric field, and a fourth arising from relativistic thermodynamic assumptions, referred to as the thermal part.   The analysis of the wormhole manifold begins in Sections~\ref{sec3} and \ref{sec4}, where we assume the existence of a massless scalar field governed solely by its kinetic energy. The main difference between these two sections lies in the separation distance of the Casimir plates, i.e., it is considered radially variable in Sec.~\ref{sec3}, whereas it is held parametrically fixed in Sec.~\ref{sec4}. In both cases, we derive an approximate analytical near-throat solution of a TW by presenting all the elements characterizing its geometry.   In Sections~\ref{sec5} and \ref{sec6}, we repeat the same procedure, but this time the scalar field possesses a potential energy  minimally coupled alongside the graviton. In all cases, the derived solutions violate the Null Energy Condition (NEC)$-$a crucial requirement for keeping the wormhole open$-$and feature a throat size that depends on the number of elementary charges the wormhole can accumulate. We summarize and conclude in Sec.~\ref{sec7}. Throughout the paper, we adopt natural units in which $\hbar = c  = 4\pi \varepsilon_0=1$ (so that $G = \ell_{\rm P}^2$), and we reintroduce physical constants when necessary.

\section{Structure of the Stress-Energy Tensor}
\label{sec2}

Consider an anisotropic fluid  described by the SET
\begin{equation}
T_{\mu \nu }=\rho (r) u_{\mu }u_{\nu }+p_{r} (r) n_{\mu }n_{\nu }+p_{t} (r) \sigma _{\mu\nu }   \label{Tmn}
\end{equation}%
where $u_{\mu }$ is the fluid four-velocity and $n_{\mu }$ is a unit
spacelike vector orthogonal to $u_{\mu }\,$, i.e., $n^{\mu }n_{\mu }=1\,$, $u^{\mu }u_{\mu }=-1\,$, $n^{\mu }u_{\mu }=0\,$. Here,
\begin{equation}
\sigma _{\mu \nu }=g_{\mu \nu }+u_{\mu }u_{\nu }-n_{\mu }n_{\nu }
\end{equation}%
is a projection operator onto a two-surface orthogonal to $u_{\mu }$ and $%
n_{\mu }\,$, i.e.,%
\begin{equation}
u_{\mu }\sigma ^{\mu \nu }\mathrm{v}_{\nu }=n_{\mu }\sigma ^{\mu \nu }%
\mathrm{v}_{\nu }=0\qquad \forall \mathrm{v}_{\nu }\,.  \label{orto}
\end{equation}%
The vector $n_{\mu }$ can be represented as%
\begin{equation}
n_{\mu }=\sqrt{\frac{1}{1-b(r)/r}}\left( 0,1,0,0\right) .
\end{equation}%
 Before proceeding, it is useful to introduce two reference lengths:
\begin{equation} \label{r1}
r_{1}^{2}=\frac{\pi ^{3}\ell_{\rm P}^{2}}{90}
\end{equation}%
and%
\begin{equation}
r_{2}^{2}=\frac{GQ^{2}}{4\pi c^{4}\varepsilon _{0}}=\left( \frac{\hbar G}{%
c^{3}}\right) \frac{Q^{2}}{4\pi \hbar c\varepsilon _{0}}=\frac{Q^{2}\ell_{\rm P}^{2}%
}{4\pi \hbar c\varepsilon _{0}} \,.
\end{equation}%
Taking into account the quantization of charge, $Q=N e$, where $e$ is the elementary charge and $N$ is an integer number, we find
\begin{equation} \label{r2}
r_{2}^{2}=  \alpha N^2 \ell_{\rm P}^{2}
\end{equation}%
with $\alpha \approx 1/137$ being the fine structure constant. If $N\leq 6\,$, then $r_1 > r_2\,$, while the opposite holds for $N \geq 7\,$. Exact equality is excluded ($r_1\neq r_2$), since it would correspond to $N \simeq 6.87\,$, which is not an integer.  To proceed with a physically meaningful source, we now assume that the energy density is given by
\begin{equation}
\rho \left( r\right) =\rho _{S}\left( r\right) +\rho _{C}\left( r\right)
+\rho _{E}\left( r\right) +\rho _{\tau }\left( r\right) =\rho _{S}\left(
r\right) -\frac{r_{1}^{2}}{\kappa r^{4}}+\frac{r_{2}^{2}}{\kappa r^{4}}+\rho
_{\tau }\left( r\right)   \label{rhoV}
\end{equation}%
where $\rho _{S}\left( r\right) $ is the scalar source, $\rho _{C}\left(
r\right) $ is the Casimir source, $\rho _{E}\left( r\right) $ is the
electric source and $\rho _{\tau }( r) $ is the thermal
energy density source. The thermal energy density is a part of
the thermal stress-energy tensor, which arises from relativistic thermodynamic considerations based on local thermal motions of particles or matter fields according to \cite{Hayward} and includes also a radial component $%
\tau _{r}(r)$ and a transverse component $\tau _{t}(r)$.  The same decomposition
will be adopted for the rest of the SET, namely%
\begin{equation}
p_{r}\left( r\right) =p_{r}^{S}\left( r\right) +p_{r}^{C}\left( r\right)
+p_{r}^{E}\left( r\right) +\tau _{r}\left( r\right) =p_{r}^{S}\left(
r\right) -\frac{3r_{1}^{2}}{\kappa r^{4}}-\frac{r_{2}^{2}}{\kappa r^{4}}%
+\tau _{r}\left( r\right)   \label{prV}
\end{equation}%
and%
\begin{equation}
p_{t}\left( r\right) =p_{t}^{S}\left( r\right) +p_{t}^{C}\left( r\right)
+p_{t}^{E}\left( r\right) +\tau _{t}\left( r\right) =p_{t}^{S}\left(
r\right) +\frac{r_{1}^{2}}{\kappa r^{4}}+\frac{r_{2}^{2}}{\kappa r^{4}}+\tau
_{t}\left( r\right) .  \label{ptV}
\end{equation}%
The scalar SET is different if we include a potential or if we exclude it.
The scalar field without a potential, hereafter referred  to as the massless scalar
field, is governed by the scalar field equation $\nabla^2\psi=0$ and is represented by the following quadruple \cite{RGAZ}
\begin{align}
 \left( \rho_{S}(r) ,p_{r}^{S}\left( r\right),p_{t}^{S}\left( r\right) , p_{t}^{S}\left( r\right) \right)  
 =\left( \frac{Ce^{-2\Phi (r)}}{2r^{4}},\frac{Ce^{-2\Phi (r)}}{2r^{4}},-%
\frac{Ce^{-2\Phi (r)}}{2r^{4}},-\frac{Ce^{-2\Phi (r)}}{2r^{4}}\right) \,,
\end{align}%
while the scalar field with a potential, hereafter referred to as the massive
scalar field, is governed by the scalar field equation $\nabla^2 \psi= \mathrm{d}V(\psi)/\mathrm{d}\psi$ and is represented by this other quadruple:
\begin{equation}
\begin{array}{c}
\rho_{S}(r)  =\frac{1}{2}\left( 1-\frac{b(r)}{r}\right) \psi ^{\prime
}(r)^{2}+V(\psi ) \\ 
p_{r}^{S} (r) =\frac{1}{2}\left( 1-\frac{b(r)}{r}\right) \psi ^{\prime
}(r)^{2}-V(\psi ) \\ 
 p_{t}^{S}(r) =-\frac{1}{2}\left( 1-\frac{b(r)}{r}\right) \psi ^{\prime
}(r)^{2}-V(\psi ) \\ 
 p_{t}^{S}(r) =-\frac{1}{2}\left( 1-\frac{b(r)}{r}\right) \psi ^{\prime
}(r)^{2}-V(\psi ) \,.%
\end{array}%
\end{equation}%
It is important to note that, in the case of a massless scalar field, the integration constant $C$ is taken to be positive in order to avoid the introduction of \textit{phantom} sources. Moreover, in all cases, we set $\rho_{\tau}(r) = 0$ for the thermal contribution.
This corresponds to a static configuration in local thermal equilibrium, where the random
microscopic motions of particles are isotropic and their averaged energy density is
negligible compared to the Casimir vacuum term. The thermal tensor, however, retains its
general form to allow for possible anisotropic pressure components. This assumption also
ensures a minimal model with fewer external parameters, where the thermal pressures can be
interpreted as a backreaction of the wormhole geometry rather than as an additional energy
source.

It should be emphasized that equations \eqref{rhoV},  \eqref{prV} and
\eqref{ptV} describe a Casimir source with a radially varying plate separation. Of course, when the plates are considered parametrically fixed at some distance $r=d\,$,
the corresponding Casimir SET simply becomes%
\begin{equation}
T^{\mu}_{ \nu }|_{\rm Casimir}=\frac{r_{1}^{2}}{\kappa d^{4}} \mathrm{diag} \left( -1,-3,1,1\right)
.  \label{CSETd}
\end{equation}
It is worth noting that in the present model the Casimir plates are treated as real physical objects, placed symmetrically with respect to the wormhole throat. Their separation is assumed to be much smaller than their curvature scale, allowing the use of the standard parallel-plate Casimir energy density as a local approximation. This configuration provides a physically motivated negative-energy source compatible with the spherical symmetry of the wormhole geometry.
 Alternatively, to ensure a consistent calculation, we adopt the configuration considered in  \cite{MTY}, in which the plates are taken to be spherical. As noted in  \cite{MTY}, this approximation  introduces an error that becomes negligible when the plates lie sufficiently close to the throat-precisely the regime relevant for our analysis.

\section{Near-throat solution for a massless scalar field  with variable Casimir plates}
\label{sec3}

\begin{figure}[h!] 
\begin{center}
\includegraphics[width=0.45
\textwidth]{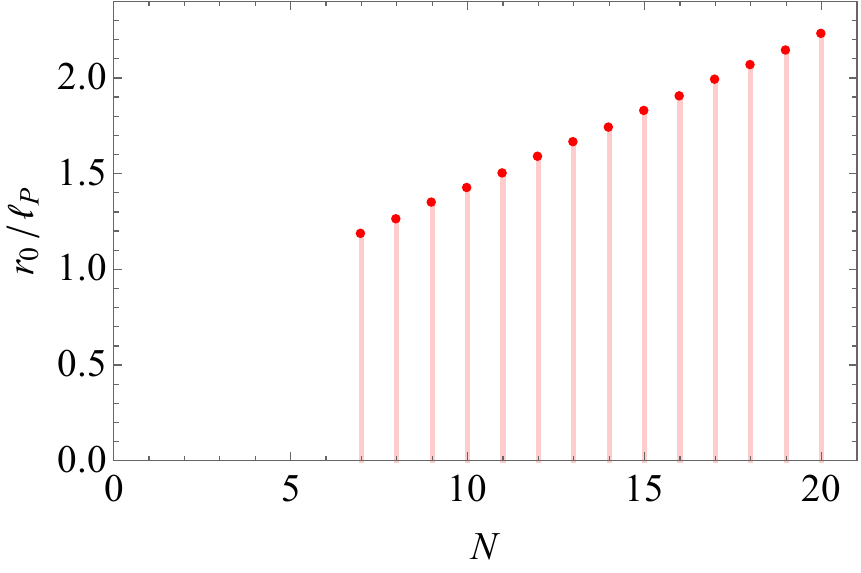}
\caption{The throat size \eqref{r0a} in Planck lengths \textit{vs} the number of elementary charges $N$.}
\label{Fig1b}
\end{center}
\end{figure}

Every TW close to the throat can be  described by a shape function approximated by the following  form
\begin{equation}
b\left(  r\right)  =r_{0}+B\left(  r-r_{0}\right)  \qquad \mathrm{with} \qquad B=b^{\prime}\left(
r_{0}\right)  <1 \,.
\label{Ab(r)}%
\end{equation}
The  function \eqref{Ab(r)}  will be used to solve the EFE in all subsequent sections. Plugging  \eqref{Ab(r)}
into  \eqref{rho}, one finds%
\begin{equation}
\Phi(r)=-\frac{1}{2}\ln\left[  \frac{2}{\kappa C}\left(  r_{1}^{2}-r_{2}%
^{2}+Br^{2}\right)  \right]  .
\label{Phi(r)S1}%
\end{equation}
Although $\Phi(r)$ seems to diverge as $r\rightarrow\infty\,$, it should be emphasized that this is a near-throat approximation. Therefore, without  loss of
generality, we can fix an  external cut-off boundary\footnote{The maximum radius $\bar{r}$ marks the cutoff of the wormhole interior, beyond which the geometry becomes asymptotically flat. In this sense, $\bar{r}$ separates the wormhole region from the external spacetime. A complete treatment of the matching between these two regions can be carried out using the Israel junction conditions \cite{Isr}, also known as the thin-shell techniques, which we leave for future work.
}
 at the location
\begin{equation}
\bar{r}=\sqrt{\frac{1}{B}\left(  \frac{\kappa C}{2}-r_{1}^{2}+r_{2}%
^{2}\right)  }\label{rb}%
\end{equation}
which is obtained by imposing $\Phi(\bar{r})=0\,$. This means that $\Phi(r)$ will be described by 
\eqref{Phi(r)S1} for $r_{0}\leq r<\bar{r}$ and $\Phi(\bar{r})=0$ for
$r\geq\bar{r}\,$. Now, we have to check the compatibility with the other EFE.
Plugging  \eqref{Ab(r)}  and   \eqref{Phi(r)S1}
in the second EFE   \eqref{pr}, one finds
\begin{equation}
\tau_{r}\!\left(  r\right)  =\frac{-2B\,r^{2}+r_{0}\left(  B-1\right)
r+2r_{1}^{2}+2r_{2}^{2}}{\kappa\,r^{4}}\label{tr(r)}%
\end{equation}
and, at the throat, we obtain%
\begin{equation}
\tau_{r}\!\left(  r_{0}\right)  =\frac{2r_{1}^{2}+2r_{2}^{2}-\left(
1+B\right)  \,r_{0}^{2}}{\kappa\,r^{4}}\,.
\end{equation}
If we impose that $\tau_{r}\!\left(  r_{0}\right)  =0\,$, we obtain a constraint on the throat size, namely
\begin{equation}
r_{0}=\sqrt{\frac{2\left(  r_{1}^{2}+r_{2}^{2}\right)  }{B+1}} \,.
\label{r0a}%
\end{equation}
The expression \eqref{r0a} can be used to rearrange 
\eqref{tr(r)} in the following manner%
\begin{equation}
\tau_{r}\!\left(  r\right)  =-\frac{\left(  2Br+Br_{0}+r_{0}\right)  \left(
r-r_{0}\right)  }{\kappa\,r^{4}} \,.\label{tr(r)1}%
\end{equation}
It is straightforward to see that $\tau_{r}\!\left(  r\right)  $ vanishes when%
\begin{equation}
r=-\frac{\left(  B+1\right)  r_{0}}{2B}\,.
\end{equation}
Such a solution can be accepted only if $-1<B<0\,$. The third EFE 
\eqref{pt} leads to%
\begin{equation}
\tau_{t}\!\left(  r\right)  =\frac{2B^{2}r^{4}+2B\left(  r_{1}^{2}-3r_{2}%
^{2}\right)  r^{2}-r_{0}\left(  r_{1}^{2}-r_{2}^{2}\right)  \left(
B-1\right)  r-4r_{1}^{2}r_{2}^{2}+4r_{2}^{4}}{2\kappa\left(  B\,r^{2}%
+r_{1}^{2}-r_{2}^{2}\right)  r^{4}}%
\end{equation}
which, at the throat, becomes%
\begin{equation}
\tau_{t}\!\left(  r_{0}\right)  =\frac{2B^{2}r_{0}^{4}+\left[  B\left(
r_{1}^{2}-5r_{2}^{2}\right)  +r_{1}^{2}-r_{2}^{2}\right]  r_{0}^{2}-4r_{1}%
^{2}r_{2}^{2}+4r_{2}^{4}}{2\kappa\left(  Br_{0}^{2}+r_{1}^{2}-r_{2}%
^{2}\right)  r_{0}^{4}} \,.
\end{equation}
Once again, we can use the constraint $\left(  \ref{r0a}\right)  $, to
rearrange $\tau_{t}\!\left(  r_{0}\right)  $ in the following manner%
\begin{equation}
\tau_{t}\!\left(  r_{0}\right)  =\frac{\left(  B-1\right)  ^{2}r_{0}%
^{4}-8r_{0}^{2}r_{1}^{2}+16r_{1}^{4}}{2\left(  r_{0}^{2}\left(  B-1\right)
+4r_{1}^{2}\right)  r_{0}^{4}\kappa} \,.
\end{equation}
Imposing $\tau_{t}\!\left(  r_{0}\right)  =0\,$, we get
\begin{equation}
r_{0}=\frac{2r_{1}}{1-B}\sqrt{1+\sqrt{B\left(  2-B\right)  }}\,\qquad \mathrm{with} \qquad
B\neq 1\,.
\label{r0b}%
\end{equation}
To have consistency we have to equate   \eqref{r0a} and
  \eqref{r0b}. We find
\begin{equation}
\sqrt{\frac{2\left(  x^{2}+1\right)  }{B+1}}=\frac{2}{1-B}\sqrt{1+\sqrt
{B\left(  2-B\right)  }}\,
\end{equation}
whose solution is%
\begin{equation}
B_{\pm}=\frac{x^{4}+4x^{2}\pm2\sqrt{2x^{6}+3x^{4}-1}-1}{x^{4}+2x^{2} +5} \label{Bpm}
\end{equation}
upon defining  $x=r_{2}/r_{1}\,$. Only $B_{-}$ can be accepted because it
satisfies the flare-out condition ($B_{-}<1$). Note also that $B\in\left[  0,1\right]  $
when $x\in\left[  1,+\infty\right)  $, meaning that $r_2>r_1$ and so $N \geq 7$. Substituting $B_-$ and $x$, along with the expressions \eqref{r1} and \eqref{r2} into \eqref{r0b}, we obtain an expression for the  throat that depends solely on the number $N$ of elementary charges. However, this expression is rather lengthy and, for this reason, we omit it here but we illustrate the $N-$dependence of the throat in Fig~\ref{Fig1b}.
Numerical analysis indicates a proportionality between the throat size and the number $N$. 
Equations  \eqref{r0a} and
\eqref{r0b} are also useful to estimate the location of the
external boundary \eqref{rb}. Indeed, by inverting the
relationships   \eqref{r0b} and   \eqref{r0a}  in
terms of $r_{2}\,$, we get
\begin{equation}
\bar{r}=\frac{\sqrt{2}\,r_{0}\sqrt{\left(  B+2\right)  \left(  1+\sqrt
{B\left(  2-B\right)  }\right)  -\left(  B-1\right)  ^{2}}}{2\sqrt{B\left(
1+\sqrt{B\left(  2-B\right)  }\right)  }}\qquad \mathrm{with} \qquad B\neq 0 \,
\label{rb1}
\end{equation}
after  assuming $C\kappa=r_{0}^{2}$ since $C$ is not fixed. When
$B\rightarrow 0\,$ then $\bar{r} \rightarrow +\infty\,$, while when
$ B\rightarrow 1$ then  $\bar{r}\rightarrow\,\sqrt{\frac{3}{2}} \,r_{0} \,$.
This limit is consistent with the approximation of being close to the throat.
The same procedure used to estimate $\bar{r}$ can also be applied to
determine whether $\tau_{t}\!\left(  \bar{r}\right)  =0\,$. We can show that as
$B\rightarrow 0\,$, $\tau_{t}\!\left(  \bar{r}\right)  \rightarrow 0\,$, while as
$B \rightarrow 1\,$, $\tau_{t}\!\left(  \bar{r}\right)  \rightarrow -2/9\,$. This
indicates that neither $\tau_{r}\!\left(  r\right)  $ nor $\tau_{t}\!\left(
r\right)  $  vanish at the boundary $\bar{r}$.

Finally, we examine the violation of the NEC, an essential requirement for the existence of a TW. Specifically, the condition $\rho+p < 0$ must be satisfied. Considering the total radial pressure along with the above expressions, the violation of the NEC is given by
\begin{equation} \label{NEC1}
  \rho(r)+ p_{r}(r) =-\frac{r_{0}\left( 1- B_-\right)
}{\kappa r^{3}}<0   
\end{equation}
which is always true.   
On the other hand,  
the quantity $\rho(r)+p_{t}(r)$ remains positive for all relevant values of $N$. Therefore, the violation of the NEC occurs exclusively through the radial pressure.

\section{Near-throat solution  for a massless scalar field with
fixed Casimir plates}
\label{sec4}

When the plates are held parametrically fixed, the Casimir SET described by \eqref{CSETd} must be considered alongside the other sources. Substituting \eqref{Ab(r)} into \eqref{rho}, we get
\begin{equation}
\Phi(r)=-\frac{1}{2}\ln\!\left(  \frac{2Br^{2}d^{4}-2d^{4}r_{2}^{2}%
+2r^{4}r_{1}^{2}}{C\,d^{4}\kappa}\right)  .\label{Phi(r)dd}%
\end{equation}
Even in this case $\Phi(r)$ appears to be divergent as $r\rightarrow\infty\,$.
However, we have to recall that this is a near-throat approximation. Therefore,
without  loss of generality, we can fix the external boundary at the location
\begin{equation}
\bar{r}=\frac{d}{2r_{1}}\sqrt{2\sqrt{B^{2}d^{4}+2C\kappa r_{1}^{2}+4r_{1}%
^{2}r_{2}^{2}}-2d^{2}B}\, \label{rbard}%
\end{equation}
with $\Phi(\bar{r})=0\,$. Accordingly, $\Phi(r)$ will be described
by \eqref{Phi(r)S1}  for $r_{0}\leq r<\bar{r}\,$ and $\Phi(r)=0$ for $r\geq\bar{r}$. Now, we have to check the compatibility with
the remaining EFE. Inserting   \eqref{Ab(r)} and \eqref{Phi(r)S1} into the second EFE  \eqref{pr}, we get
\begin{equation}
\tau_{r}\!\left(  r\right)  =\frac{\left(  -2B\,r^{2}+r_{0}\left(  B-1\right)
r+2r_{2}^{2}\right)  d^{4}+2r^{4}r_{1}^{2}}{\kappa\,d^{4}r^{4}}\label{tau(r)d}%
\end{equation}
and, at the throat, we obtain%
\begin{equation}
\tau_{r}\!\left(  r_{0}\right)  =\frac{2r_{0}^{4}r_{1}^{2}+\left(  2r_{2}%
^{2}-r_{0}^{2}\left(  B+1\right)  \right)  d^{4}}{\kappa\,d^{4}r_{0}^{4}} \,.
\end{equation}
By imposing $\tau_{r}\!\left(  r_{0}\right)  =0\,$, one finds 
\begin{equation}
r_{0}=\pm\frac{d}{2r_{1}}\sqrt{d^{2}B+d^{2}\pm\sqrt{B^{2}d^{4}+2B\,d^{4}%
+d^{4}-16r_{1}^{2}r_{2}^{2}}} \,.\label{r0d}%
\end{equation}
Only the full positive solution can be accepted. It is straightforward to see
that if $r_{2}$ becomes too large, $r_{0}$ becomes imaginary. To avoid this, we additionally impose that
\begin{equation}
d^{4}-16r_{1}^{2}r_{2}^{2}=0 \,.
\end{equation}
This assumption also establishes a correspondence between the plate separation and $r_{2}\,$, i.e., 
\begin{equation}
d= 2\sqrt{ r_{1}r_{2}} \,.\label{d}
\end{equation}
 Substituting \eqref{d} into \eqref{r0d}, we obtain
\begin{equation}
r_{0}=2r_{2}\sqrt{B+1+\sqrt{B^{2}+2B}} \,.\label{r0dd}
\end{equation}
 Equations   \eqref{d} and   \eqref{r0dd} are useful for rearranging the profile of $\tau_{t}\!\left(  r\right) \, $. Indeed, we can write
\begin{equation}
\tau_{t}\!\left(  r\right)  =\frac{2B\,r^{6}r_{1}^{2}+r_{0}r_{1}^{2}\left(
B-1\right)  r^{5}+\left(  2B^{2}d^{4}-4r_{1}^{2}r_{2}^{2}\right)
r^{4}-6B\,d^{4}r^{2}r_{2}^{2}+d^{4}r_{0}r_{2}^{2}\left(  B-1\right)
r+4d^{4}r_{2}^{4}}{2\left(  B\,r^{2}d^{4}-r_{2}^{2}d^{4}+r_{1}^{2}%
r^{4}\right)  r^{4}\kappa}%
\end{equation}
which, at the throat, becomes
\begin{equation}
\tau_{t}\!\left(  r_{0}\right)  =\frac{r_{1}^{2}\left(  3B-1\right)  r_{0}%
^{6}+\left(  2B^{2}d^{4}-4r_{1}^{2}r_{2}^{2}\right)  r_{0}^{4}-d^{4}\left(
5B+1\right)  r_{2}^{2}r_{0}^{2}+4d^{4}r_{2}^{4}}{2\left(  B\,d^{4}r_{0}%
^{2}-r_{2}^{2}d^{4}+r_{1}^{2}r_{0}^{4}\right)  r_{0}^{4}\kappa}%
\end{equation}
and finally%
\begin{equation}
\tau_{t}\!\left(  r_{0}\right)  =\frac{\left(  14B^{3}+18B^{2}-3B-3\right)
\sqrt{B}\sqrt{B+2}+14B^{4}+32B^{3}+8B^{2}-8B-1}{\left(  2\left(  3B+1\right)
\sqrt{B}\sqrt{B+2}+6B^{2}+8B\right)  \kappa\left(  1+B+\sqrt{B}\,\sqrt
{B+2}\right)  r_{0}^{2}} \,.
\end{equation}
From the condition $\tau_t(r_0)=0\,$, we obtain  the numerical value%
\begin{equation}
B=0.425 \label{Bn}%
\end{equation}
and  the throat size becomes
\begin{equation}
r_0 \approx 0.27 N \ell_{\rm P}
\end{equation}
indicating a proportionality with the number of elementary charges.
With the help of  \eqref{d} and    \eqref{r0dd}, we can rearrange $\tau_{r}\!\left(  r\right)  $ in the following
form
\begin{equation}
\tau_{r}\!\left(  r\right)  =\left(  r-r_{0}\right)  \frac{2r\left(
r^{2}+r_{0}r-r_{0}^{2}\right)  \left(  B+1\right)  \sqrt{B\left(  B+2\right)
}+\left(  2B^{2}+4B+1\right)  \left(  r^{3}+r^{2}r_{0}\right)  +\left(
1-2B^{2}\right)  r_{0}^{2}r-r_{0}^{3}}{2r^{4}\kappa r_{0}^{2}\left(
1+B+\sqrt{B\left(  B+2\right)  }\right)  }\,.
\end{equation}
The component  $\tau_{r}\!\left(  r\right)  $  does not vanish at any point except at the throat. In terms
of $r_{0}\,$, the component $\tau_{t}\!\left(  r\right)  $ can be expressed as%
\begin{equation}
\tau_{t}\!\left(  r\right)  =\frac{A^{3}\left(  B\right)  \left(
2B\,r^{6}+\left(  B-1\right)  r_{0}r^{5}\right)  +A^{2}\left(  B\right)
\left(  8B^{2}-1\right)  r_{0}^{2}r^{4}-A\left(  B\right)  r_{0}^{4}r\left(
6Br+r_{0}\left(  B-1\right)  \right)  +r_{0}^{6}}{2\left(  A\left(  B\right)
4Br_{0}^{2}r^{2}+A^{2}\left(  B\right)  r^{4}-r_{0}^{4}\right)  r^{4}\kappa
A\left(  B\right)  } 
\end{equation}
where we have defined%
\begin{equation}
A\left(  B\right)  =\sqrt{B\left(  B+2\right)  }+B+1 \,.
\end{equation}
Also in this case, we cannot find a point where $\tau_{t}\!\left(  r\right)  $
vanishes. To verify whether $\bar{r}$ is compatible with the EFE, we note that there remains an unfixed parameter, namely the value of $C$. Using  \eqref{d} and \eqref{r0dd}, the boundary $\bar{r}$ becomes
\begin{equation}
\bar{r}=\frac{\sqrt{\sqrt{2C\kappa A\left(  B\right)  +4B^{2}r_{0}^{2}%
+r_{0}^{2}}-2Br_{0}}}{\sqrt{A\left(  B\right)  }} \sqrt{r_{0}} \,.
\end{equation}
A compatible boundary must satisfy $\bar{r}>$ $r_{0}\,$. By  defining the dimensionless constant
\begin{equation}
x=\frac{C\kappa}{r_{0}^{2}} \,
\end{equation}
one gets%
\begin{equation}
\bar{r}=\frac{\sqrt{\sqrt{2xA\left(  B\right)  +4B^{2}+1}-2B}}
{\sqrt{A\left(  B\right)  }}r_{0}
\end{equation}
and by plugging the value  $B=0.425$ inside $\bar{r}\,$, one
finds that $\bar{r}>r_{0}$ if and only if $x>1.864\,$.

Last but not least, by considering the total radial pressure, the NEC takes the form
\begin{equation}
\rho(r)+p_{r}(r) = - \frac{(1-B)r_0}{\kappa r^3} 
\end{equation}
and  is always violated for $B=0.425\,$, while numerical analysis shows that by using the tangential pressure, we get always $\rho(r)+p_{t}(r)>0\,$.

\section{Near-throat solution for a massive scalar field with variable Casimir plates}
\label{sec5}

When a potential is included as an additional source, the scalar field
equation becomes
\begin{equation}
\left(  1-\frac{b(r)}{r}\right)  \left[  \psi^{\prime\prime}(r)+\left(
\frac{2}{r}+\Phi^{\prime}(r)\right)  \psi^{\prime}(r)\right]  +\left(
\frac{b(r)-rb^{\prime}(r)}{2r^{2}}\right)  \psi^{\prime}(r)=\frac
{\mathrm{d}V(\psi)}{\mathrm{d}\psi}\,. \label{scalar_eom}%
\end{equation}
Since the equation of motion is considerably complicated with respect to the
massless case, such a complication is also transported to the SET, represented
by
\begin{align}
\rho\left(  r\right)   &  =\frac{1}{2}\left(  1-\frac{b(r)}{r}\right)
\psi^{\prime}(r)^{2}+V(\psi)-\frac{r_{1}^{2}}{\kappa r^{4}}+\frac{r_{2}^{2}%
}{\kappa r^{4}}  \\
p_{r}\left(  r\right)   &  =\frac{1}{2}\left(  1-\frac{b(r)}{r}\right)
\psi^{\prime}(r)^{2}-V(\psi)-\frac{3r_{1}^{2}}{\kappa r^{4}}-\frac{r_{2}^{2}%
}{\kappa r^{4}}+\tau_{r}\left(  r\right)  \\
p_{t}\left(  r\right)   &  =-\frac{1}{2}\left(  1-\frac{b(r)}{r}\right)
\psi^{\prime}(r)^{2}-V(\psi)+\frac{r_{1}^{2}}{\kappa r^{4}}+\frac{r_{2}^{2}%
}{\kappa r^{4}}+\tau_{t}\left(  r\right)  .
\end{align}
However, if we adopt the approximated profile of the shape function $\left(
\ref{Ab(r)}\right)  $, things are different because the SET reduces to%
\begin{align}
\rho\left(  r\right)   &  =V(\psi)-\frac{r_{1}^{2}}{\kappa r^{4}}+\frac
{r_{2}^{2}}{\kappa r^{4}}  \label{rhop}\\
p_{r}\left(  r\right)   &  =-V(\psi)-\frac{3r_{1}^{2}}{\kappa r^{4}}%
-\frac{r_{2}^{2}}{\kappa r^{4}}+\tau_{r}\left(  r\right)  \label{prp}\\
p_{t}\left(  r\right)   &  =-V(\psi)+\frac{r_{1}^{2}}{\kappa r^{4}}%
+\frac{r_{2}^{2}}{\kappa r^{4}}+\tau_{t}\left(  r\right)  . \label{ptp}%
\end{align}
Moreover, the scalar field equation $\left(  \ref{scalar_eom}\right)  $ can be
approximated with%
\begin{equation}
\left(  \frac{b(r)-rb^{\prime}(r)}{2r^{2}}\right)  \psi^{\prime}%
(r)=\frac{\mathrm{d}V(\psi)}{\mathrm{d}\psi}=\frac{V^{\prime}(r)}{\psi
^{\prime}(r)}  \label{appMS}%
\end{equation}
 leading to%
\begin{equation}
\psi^{\prime}(r)=r\sqrt{\frac{2V^{\prime}(r)}{b(r)-rb^{\prime}(r)}} \,.
\label{EoMA}
\end{equation}
Since the scalar field depends only on the radial coordinate, $\psi=\psi(r)\,$, the potential can be expressed as a function of $r\,$, i.e., $V(\psi)=V(r)\,$. It is interesting to note that, in this framework, the NEC  becomes
\begin{align}
\rho\left(  r\right)  +p_{r}\left(  r\right)   &  =-\frac{4r_{1}^{2}}{\kappa r^{4}}  +\tau
_{r}\left(  r\right)  . \label{NEC}%
\end{align}
This means that the role of the thermal tensor is crucial to establish if
the NEC can be violated or not. To obtain this information, we begin by solving the first EFE \eqref{rho}, which yields
\begin{equation}
\frac{B}{r^{2}}=\kappa V(r)-\frac{r_{1}^{2}}{r^{4}}+\frac{r_{2}^{2}}{r^{4}%
}\,.
\end{equation}
 Solving
with respect to the potential, we can get its analytical representation in
terms of $r$, namely%
\begin{equation}
V(r)=\frac{B}{\kappa r^{2}}+\frac{r_{1}^{2}}{\kappa r^{4}}-\frac{r_{2}^{2}%
}{\kappa r^{4}} \,. \label{V(P)}%
\end{equation}
The second EFE \eqref{pr}, within the same near-throat approximation and using \eqref{V(P)},   can be rewritten in terms of $\tau_{r}\!\left(  r\right)  $ as
\begin{equation}
\tau_{r}\left(  r\right)  =\frac{4r_{1}^{2}}{\kappa r^{4}}+\frac{r_{0}\left(  B-1\right)  }{\kappa
r^{3}} \,. \label{taur}%
\end{equation}
Finally, from the
third EFE \eqref{pt}, we get
\begin{align}
\tau_{t}\left(  r\right)   =  \frac{r_{0}\left(  1-B\right)  }{2\kappa r^{2}}\left(  \Phi\!^{\prime
}\left(  r\right)  +\frac{1}{r}\right)  +\frac{B}{\kappa r^{2}}-\frac
{2r_{2}^{2}}{\kappa r^{4}} \,. \label{tautP}%
\end{align}
To further proceed, we need to use the SET conservation equation represented by%

\begin{gather}
\left(  1-\frac{b(r)}{r}\right)  \left[  \psi^{\prime\prime}(r)\psi^{\prime
}(r)+\left(  \frac{2}{r}+\Phi\!^{\prime}\left(  r\right)  \right)
\psi^{\prime}(r)^{2}\right]  +\left(  \frac{b(r)-rb^{\prime}(r)}{2r^{2}%
}\right)  \psi^{\prime}(r)^{2}-V^{\prime}(r)\nonumber\\
=\frac{\left(  \tau_{r}\!\left(  r\right)  \kappa\,r^{5}-4rr_{1}^{2}\right)
\Phi^{\prime}\!\left(  r\right)  +\tau_{r}^{\prime}\!\left(  r\right)
\kappa\,r^{5}+2\tau_{r}\!\left(  r\right)  \kappa\,r^{4}-2\tau_{t}\!\left(
r\right)  \kappa\,r^{4}+4r_{1}^{2}}{\kappa\,r^{5}}\,. \label{SETM}%
\end{gather}
The first term of the above equation is identically zero because it coincides with the equation of motion for the scalar field \eqref{scalar_eom}, giving
\begin{equation}
\frac{\left(  \tau_{r}\!\left(  r\right)  \kappa\,r^{5}-4rr_{1}^{2}\right)
\Phi^{\prime}\!\left(  r\right)  +\tau_{r}^{\prime}\!\left(  r\right)
\kappa\,r^{5}+2\tau_{r}\!\left(  r\right)  \kappa\,r^{4}-2\tau_{t}\!\left(
r\right)  \kappa\,r^{4}+4r_{1}^{2}}{\kappa\,r^{5}}\,=0 \,.\label{SETM1}
\end{equation}
This implies that the scalar field contribution is irrelevant in satisfying the SET conservation equation. To determine the profile of the redshift function, we simplify \eqref{SETM1} using \eqref{taur} and \eqref{tautP}. This yields
\begin{equation}
\Phi ^{\prime} (  r )  =\frac{B\,r^{2}+2r_{1}^{2}-2r_{2}^{2}}%
{r^{2}r_{0}\left(  B-1\right)  }
\end{equation}
whose solution is
\begin{equation}
\Phi\!\left(  r\right)  =\frac{r^{2}B-2r_{1}^{2}+2r_{2}^{2}}{rr_{0}\left(
B-1\right)  }+K \, \label{Phi(r)M}%
\end{equation}
with $K$ being an integration constant. Since the location of the external boundary of such an approximation is
unknown, we may tempt to fix the external boundary $\bar{r}$ in such a way that
$
\Phi\!\left(  \bar{r}\right)  =0\,$. Such a condition specifies the integration constant $K$ to be
\begin{equation}
    K= -\frac{\bar{r}^{2}B-2r_{1}^{2}+2r_{2}^{2}}{\bar{r}r_{0}\left(
B-1\right) }\,,
\end{equation}
leading to
\begin{equation}
\Phi \!\left( r\right) =\frac{r^{2}B-2r_{1}^{2}+2r_{2}^{2}}{rr_{0}\left(
B-1\right) }-\frac{\bar{r}^{2}B-2r_{1}^{2}+2r_{2}^{2}}{\bar{r}r_{0}\left(
B-1\right) }\,.\label{Phi(r)Mb}
\end{equation}
Next, by plugging \eqref{Phi(r)Mb} into\eqref{tautP}, 
one obtains
\begin{equation}
\tau_{t}\!\left(  r\right)  =\frac{B\,r^{2}-2r_{1}^{2}-2r_{2}^{2}-r_{0}\left(
B-1\right)  r}{2r^{4}\kappa} \,.
\end{equation}
Since the role of the thermal tensor is to balance the EFE, it is natural to consider minimizing its contribution at the throat. To this end, we impose
\begin{equation}
\tau_{t}\!\left(  r_{0}\right)  =\frac{r_{0}^{2}-2r_{1}^{2}-2r_{2}^{2}}{2r_{0}^{4}\kappa}=0
\end{equation}
which yields a throat size that depends explicitly on the number of charges:
\begin{equation} \label{r0V}
r_{0}=\sqrt{2\left(  r_{1}^{2}+r_{2}^{2}\right)  } = \ell_{\rm P} \sqrt{\frac{\pi^3}{45}+\frac{2N^2}{137}}\,
\end{equation}
after using  the relations \eqref{r1} and \eqref{r2}.
With the help of \eqref{r0V}, the component $\tau_{t}\!\left(  r\right)
$ can be rearranged in the following way%
\begin{equation}
\tau_{t}\!\left(  r\right) =\frac{\left(  Br+r_{0}\right)  \,\left(  r-r_{0}\right)
}{2r^{4}\kappa} \,.
\end{equation}
At the other hand, if we apply again the same constraint, namely
\begin{equation}
\tau_{r}\left(  r_{0}\right)  =\frac{1}{\kappa r_{0}^{4}}\left(  4r_{1}%
^{2}+r_{0}^{2}\left(  B-1\right)  \right)  =0 \,,
\end{equation}
we obtain
\begin{equation}
r_{0}=\frac{2r_{1}}{\sqrt{1-B}} \,. \label{r01}
\end{equation}
Now with the help of   \eqref{r01}, we can
rearrange $\tau_{r}\left(  r\right)  $ in the following way%
\begin{equation}
\tau_{r}\left(  r\right)  =\frac{r_{0}\left(  B-1\right)  }{\kappa r^{4}}\left(  r-r_{0}\right)  .
\end{equation}
Note that the compatibility of the two values of the throat is possible if%
\begin{equation}
B=\frac{r_{2}^{2}-r_{1}^{2}}{r_{2}^{2}+r_{1}^{2}} \,. \label{Bvp}
\end{equation}
As we can see, the constant $B$ is always less than one, thereby satisfying the flare-out condition. It is also important to note that the thermal pressures $\tau_{r}\left(  r\right)  $ and
$\tau_{t}\left(  r\right)  $ vanish only at the throat. There exists no
external boundary at which $\tau_{r}\left(  r\right)  $ and $\tau_{t}\left(
r\right)  $ can be set to zero.
Since it is important to find a boundary, we adopt a different strategy.
Instead of searching for a vanishing point,  we determine the location where one component of the thermal tensor has at least one stationary point. Examining $\tau_{r} (  r ) $, it is clear that the derivative $\tau_{r}^{\prime}\left(  \bar{r}\right)$ vanishes at
\begin{equation} \label{rbar1}
\bar{r}=\frac{4}{3}r_{0} \,.
\end{equation}
This means that%
\begin{equation}
\tau_{r}\left(  \bar{r}\right)  =\frac{27\left(  B-1\right)  }{256\kappa
r_{0}^{2}} \,.\label{trrb} 
\end{equation}
This value remains very small as long as $r_{2}\gg r_{1}$. On the other hand,
if we apply the same procedure to $\tau_{t}\left(  r\right)  $, we find
\begin{equation}
\tau_{t}^{\prime}\left(  r\right)  =\frac{-2B\,r^{2}+3rr_{0}\left(
B-1\right)  +4r_{0}^{2}}{2r^{5}\kappa}=0\qquad \Longleftrightarrow \qquad \bar{r}=\frac{\left(  3\left(  B-1\right)  \pm\sqrt{9B^{2}+14B+9}\right)  r_{0}}{4B} \,.
\label{rbar2}
\end{equation}
Only the positive solution can be accepted. When $B\in\left(  0,1\right)  $,
$\bar{r}\in\left(  4/3r_{0},\sqrt{2}r_{0}\right)  $ and%
\begin{equation}
\tau_{t}\left(  \bar{r}\right)  =-\frac{8\left(  B+3-\sqrt{9B^{2}%
+14B+9}\right)  \left(  3B+1+\sqrt{9B^{2}+14B+9}\right)  B^{3}}{r_{0}%
^{2}\kappa\left(  3\left(  B-1\right)  +\sqrt{9B^{2}+14B+9}\right)  ^{4}}\,.
\end{equation}
When $B=0$ then $\tau_{t}\left(  \bar{r}\right)  =0\,$. However, $B=0$ must be
discarded. On the other hand, when $B\rightarrow 1$ then
\begin{equation}
\tau_{t}\left(  \bar{r}\right)  =\frac{1}{8r_{0}^{2}\kappa}\,.
\end{equation}
For both choices of $\bar{r}$, namely equations \eqref{rbar1} and \eqref{rbar2}, the contribution of the thermal tensor is
small. Nevertheless, due to the dependence on $B$ in  \eqref{trrb}, it is easy to see that $\tau_{r}\left(  \bar{r}\right)  \rightarrow 0$ when $B\rightarrow 1$. 

However, a different situation arises when considering the total transverse pressure. From \eqref{ptp}, we find
\begin{gather}
p_{t}\left(  r\right)  =\frac{-Br^{2}-r_{0}\left(  B-1\right)  r-r_{0}^{2}}{2\kappa r^{4}}%
\end{gather}
and, by implying $p_{t}\left(  \bar{r}\right)=0$, we conclude to
\begin{equation}
\bar{r}=\frac{r_{0}}{2B}\left(  \sqrt{5B^{2}-2B+1}+1-B\right)  . \label{rbar}%
\end{equation}
If the boundary $\bar{r}$ is close to $r_{0}\,$, then $B$ has to be
close to unity. This is guaranteed by  \eqref{Bvp}. The boundary $\bar{r}$
represents the point where $p_{t}\left(  r\right)  $ changes sign. It is important to emphasize that neither the radial pressure
$p_{r}\left(  r\right)  $ nor the energy density $\rho\left(  r\right)  $ exhibit this behaviour. Therefore, we can formally regard the expression in \eqref{rbar} as defining the boundary $\bar{r}$ of the TW.

As a final step, the NEC is violated through
\begin{equation}
\rho(r)+p_r(r) = - \frac{4r_1^2}{\kappa r_0^2 r^3}<0
\end{equation}
as well as through the tangential pressure, $\rho(r)+p_t(r)<0\,$, as long as $N \leq 6$ for the second expression. Nevertheless, this range for the number of electric charges implies $r_2<r_1\,$, while for  $N \geq 7$ ($r_2>r_1$), we get  $\rho(r)+p_t(r)>0\,$.

\section{Near-throat solution for a massive scalar field with fixed Casimir plates}
\label{sec6}

In this section, we will consider the same source of the previous section but
with the Casimir plates treated as parametrically fixed. The new SET, within the same near-throat approximation, is given by
\begin{align}
\rho\left(  r\right)   &  =V(r)-\frac{r_{1}^{2}}{\kappa d^{4}}+\frac
{r_{2}^{2}}{\kappa r^{4}} \\
p_{r}\left(  r\right)   &  =-V(r)-\frac{3r_{1}^{2}}{\kappa d^{4}}%
-\frac{r_{2}^{2}}{\kappa r^{4}}+\tau_{r}\left(  r\right)   \\
p_{t}\left(  r\right)   &  =-V(r)+\frac{r_{1}^{2}}{\kappa d^{4}}%
+\frac{r_{2}^{2}}{\kappa r^{4}}+\tau_{t}\left(  r\right)  
\end{align}
and the first EFE  reduces to
\begin{equation}
\frac{B}{r^{2}}=\kappa V(r)-\frac{r_{1}^{2}}{d^{4}}+\frac{r_{2}^{2}}{r^{4}%
} \,
\end{equation}
after considering the approximated profile \eqref{Ab(r)}. Solving with respect to the potential, we obtain
\begin{equation}
V(r)=\frac{B}{\kappa r^{2}}+\frac{r_{1}^{2}}{\kappa d^{4}}-\frac{r_{2}^{2}%
}{\kappa r^{4}}\,.
\end{equation}
Using the above expression, we derive from the second EFE the form of the radial pressure
\begin{equation}
\tau_{r}\left(  r\right) 
=\frac{4r_{1}^{2}}{\kappa d^{4}}+\frac{r_{0}\left(  B-1\right)  }{\kappa
r^{3}} \,.\label{taurd}
\end{equation}
The third EFE gives us
\begin{equation}
\tau_{t} (  r)    =\left(\frac{b\left(  r\right)  -b^{\prime}\left( r\right)  r}{2\kappa r^{2}}+\frac{r_{0}\left(  1-B\right)  }{2\kappa r^{2}}\right)\left(  \Phi^{\prime}(  r)  +\frac
{1}{r}\right) +\frac{2B}{\kappa r^{2}}-\frac
{4r_{2}^{2}}{\kappa r^{4}}   \,. \label{tautPd}%
\end{equation}
It is interesting to note that the third EFE does not develop a dependence on
the plates separation $d$. Following the previous section, we will use the SET
conservation which, in this case, is described by%
\begin{gather}
\left(  1-\frac{b(r)}{r}\right)  \left[  \psi^{\prime\prime}(r)\psi^{\prime
}(r)+\left(  \frac{2}{r}+\Phi\!^{\prime}\left(  r\right)  \right)
\psi^{\prime}(r)^{2}\right]  +\left(  \frac{b(r)-rb^{\prime}(r)}{2r^{2}%
}\right)  \psi^{\prime}(r)^{2}-V^{\prime}(r)\nonumber\\
=\frac{r\left(  \tau_{r}\!\left(  r\right)  \kappa\,d^{4}-4r_{1}^{2}\right)
\Phi^{\prime}\!\left(  r\right)  +\tau_{r}^{\prime}\!\left(  r\right)  \kappa
rd^{4}+2\tau_{r}\!\left(  r\right)  \kappa d^{4}-2\tau_{t}\!\left(  r\right)
\kappa d^{4}-8r_{1}^{2}}{\kappa r\,d^{4}}\,.
\end{gather}
Close to the throat, we can write%
\begin{gather}
\left(  \frac{b(r)-rb^{\prime}(r)}{2r^{2}}\right)  \psi^{\prime}%
(r)^{2}-V^{\prime}(r)\nonumber\\
=\frac{r\left(  \tau_{r}\!\left(  r\right)  \kappa\,d^{4}-4r_{1}^{2}\right)
\Phi^{\prime}\!\left(  r\right)  +\tau_{r}^{\prime}\!\left(  r\right)  \kappa
rd^{4}+2\tau_{r}\!\left(  r\right)  \kappa d^{4}-2\tau_{t}\!\left(  r\right)
\kappa d^{4}-8r_{1}^{2}}{\kappa r\,d^{4}}\,\label{SET3B}%
\end{gather}
and with the help of   \eqref{EoMA}, one finds that
\eqref{SET3B}   becomes%
\begin{equation} \label{dfgh}
\frac{r\left(  \tau_{r}\!\left(  r\right)  \kappa\,d^{4}-4r_{1}^{2}\right)
\Phi^{\prime}\!\left(  r\right)  +\tau_{r}^{\prime}\!\left(  r\right)  \kappa
rd^{4}+2\tau_{r}\!\left(  r\right)  \kappa d^{4}-2\tau_{t}\!\left(  r\right)
\kappa d^{4}-8r_{1}^{2}}{\kappa r\,d^{4}}\,=0 \,.
\end{equation}
Employing equations \eqref{taurd}  and \eqref{tautPd},  from \eqref{dfgh} we get
\begin{equation}
\Phi^{\prime} (  r )  =\frac{B\,r^{2}-2r_{2}^{2}}{r^{2}r_{0}\left( B-1\right)  } \label{Phi(r)'}%
\end{equation}
whose solution is%
\begin{equation}
\Phi\!\left(  r\right)  =\frac{r^{2}B+2r_{2}^{2}}{rr_{0}\left(  B-1\right)
}+K \,\label{Phi(r)d}%
\end{equation}
with $K$ being an integration constant. Also for this case, if we impose that 
$\Phi \!\left( \bar{r}\right) =0$ with $\bar{r}>r_{0}$, then one fixes $K$ accordingly to obtain
\begin{equation}
\Phi \!\left( r\right) =\frac{r^{2}B+2r_{2}^{2}}{rr_{0}\left( B-1\right) }-
\frac{\bar{r}^{2}B+2r_{2}^{2}}{\bar{r}r_{0}\left( B-1\right) }\,.
\label{Phi(r)db}
\end{equation} Plugging the approximated redshift function  \eqref{Phi(r)db}  
into  \eqref{tautPd}, we obtain
\begin{equation}
\tau_{t}\!\left(  r\right)  =\frac{B\,r^{2}-2r_{2}^{2}-r_{0}\left( B-1\right)  r}{2r^{4}\kappa} \,.\label{taut(r)d}
\end{equation}
Even in the case of  fixed plates, we desire to minimize the effect of the thermal
field, at least at the throat. For this purpose, we impose the condition
\begin{equation}
\tau_{t}\!\left(  r_{0}\right) =\frac{r_{0}^{2}-2r_{2}^{2}}%
{2\kappa r_{0}^{4}}=0
\end{equation}
which is satisfied if%
\begin{equation}
r_{0}=\sqrt{2}r_{2} =\ell_{\rm P} N \sqrt{\frac{2}{137}}\,.
\label{r0d1}%
\end{equation}
The throat size is again proportional to the charge number  $N$. With the help of  \eqref{r0d1}, we can rearrange $\tau
_{t}\!\left(  r\right)  $ in the following manner
\begin{equation}
\tau_{t}\!\left(  r\right)  =\frac{\left(  \,r-r_{0}\right)  \left(  Br+r_{0}\right)
}{2r^{4}\kappa} \,.
\end{equation}
The component $\tau_{t}\!\left(  r\right)  $ vanishes not only at the throat but also at
\begin{equation}
r=-\frac{r_{0}}{B} \,.
\end{equation}
Unfortunately, this can occur  only if $B$ is negative. By applying the same procedure on
$\tau_{r}\!\left(  r\right)  $, one gets%
\begin{equation}
B\,r_0^{2}-2r_{2}^{2}=\frac{d^{2}\sqrt{1-B}}{2r_{1}} \,.\label{r0d2}%
\end{equation}
Using \eqref{r0d2}, we can rewrite $\tau_{r}\!\left(  r\right)  $ as
\begin{equation}
\tau_{r}\left(  r\right)  =\frac{\left(  1-B\right)  \left(  r^{3}-r_{0}%
^{3}\right)  }{\kappa r^{3}r_{0}^{2}}%
\end{equation}
which shows that it vanishes only at the throat. To proceed further, we need to verify whether the value of $r_{0}$ given in \eqref{r0d1} is consistent with that in	  \eqref{r0d2}. This can be achieved by solving for $B$. The solution is
\begin{equation}
B=1-\frac{8r_{2}^{2}r_{1}^{2}}{d^{4}}\,\label{B}
\end{equation}
satisfying always the flare-out condition. If we also impose $B>0\,$, we get 
\begin{equation}
d>\sqrt{2\sqrt{2}r_{2}r_{1}} \,.
\end{equation}
Since $d$ is fixed by the physical setup, this imposes a constraint on $r_{2}$ and, consequently, on the maximum number of charges the TW can accumulate:
\begin{equation}
 N < 12.5 \left(\frac{d}{\ell_{\rm P}}\right)^2 .
\end{equation}
Since  $B$ cannot be negative,
we conclude that neither $\tau_{t}\!\left(  r\right)  $ nor $\tau_{r}\!\left(
r\right)  $ admit an external boundary where they vanish.  Nevertheless, observing \eqref{Phi(r)'},  one sees that there exists a point where $\Phi ^{\prime} (  \bar{r} )  =0\,$. This is located at
\begin{equation}
\bar{r}= r_{2} \sqrt{\frac{2}{B}}=\frac{r_{0}}{\sqrt{B}} \,.
\end{equation}
Since $0<B<1\,$, it follows that  $\bar{r}>r_{0}\,$. Therefore, the TW can be consistently defined within the interval
$\left[  r_{0},\bar{r}\right]  \,$.
On the other hand, it is possible to adopt
the procedure from the previous section by searching for stationary points of
$\tau_{r}\left(  r\right)  $ and $\tau_{t}\left(  r\right)  $. It is immediately evident that $\tau_{r}\left(  r\right)  $ has no stationary points, while $\tau_{t}\left(  r\right)  $ exhibits the same stationary point identified earlier, thus leading to the same conclusion. In other words, an external boundary can be located at
\begin{equation}
\bar{r}=\frac{\left(  3\left(  B-1\right)  +\sqrt{9B^{2}+14B+9}\right)  r_{0}}{4B} 
\end{equation}
with%
\begin{equation}
\tau_{t}\left(  \bar{r}\right)  =\frac{1}{8r_{0}^{2}\kappa}%
\end{equation}
since only the values of $B$ close to $1$ can be accepted.

Finally, we check the violation of the NEC. For the radial pressure,
\begin{equation}
\rho(r)+p_{r}(r) = - \frac{2r_0 r_1^2}{\kappa r^3(r_1^2+r_2^2)} <0 \,,
\end{equation}
the NEC is always violated. In the case of the tangential pressure, we find that $\rho(r) + p_{t}(r) < 0$ only if $N \leq 3\,$, which sets an upper bound on the throat size at $N = 3$ (corresponding to $r_0 \approx 0.36 \ell_{\rm P}$). However, it is not necessary to impose such a restrictive sub-Planckian bound on the throat, since the NEC is already violated via the radial pressure, allowing us to set $N \geq 4\,$.

\section{Conclusions}
\label{sec7}

In this paper we have explored how various physical sources can contribute to the sustainability of a traversable wormhole. Specifically, we have examined the combined effects of a scalar field (with and without a potential), a static electric field, and a Casimir source. This combination of sources generates a stress-energy tensor that significantly complicates the search for solutions to the Einstein field equations. To address this, we have restricted our analysis to a region near the throat. This approximation allows for a simplified form of the shape function, given by \eqref{Ab(r)}. Despite this simplification, it should be noted that any TW can be represented by \eqref{Ab(r)} close to the throat. The only
difference between the various profiles is encoded in the coefficient $B$. 
 Once the form of the shape function was fixed, we divided our investigation into four parts; two without a potential for the scalar field and two with a potential. In both scenarios, we further distinguished the analysis by considering the Casimir source under two configurations; one where the plates vary radially  and another where the plates are held parametrically fixed. For all cases, the introduction of an additional unknown tensor$-$the \textit{thermal tensor}$-$has been necessary to ensure consistency in the Einstein field equations. This tensor does not contribute to the energy density. It affects only the spatial pressure components and plays a fundamental role  in the discussion of the null energy condition. Indeed, for the massless case, one finds that
\begin{equation}
\rho (  r )  +p_{r} (  r )  =\left\{
\begin{array}
[c]{c}%
\frac{Ce^{-2\Phi(r)}}{r^{4}}-\frac{4r_{1}^{2}}{\kappa r^{4}}+\tau_{r}\left(
r\right)  \\
\\
\frac{Ce^{-2\Phi(r)}}{r^{4}}-\frac{4r_{1}^{2}}{\kappa d^{4}}+\tau_{r}\left(
r\right)
\end{array}
\right.  ,
\end{equation}
while for the massive case, one finds%
\begin{equation}
\rho (  r )  +p_{r} (  r )  =\left\{
\begin{array}
[c]{c}%
-\frac{4r_{1}^{2}}{\kappa r^{4}}+\tau_{r}\left(  r\right)  \\
\\
-\frac{4r_{1}^{2}}{\kappa d^{4}}+\tau_{r}\left(  r\right)
\end{array}
\right.  .
\end{equation}
It is interesting to note that in all four cases
\begin{equation} \label{NEC_final}
\rho\left(  r\right)  +p_{r}\left(  r\right)  =-\frac{r_{0}\left(  1-B\right)
}{\kappa r^{3}}<0
\end{equation}
due to the flare-out condition. Therefore, the NEC is always violated. 
A convenient way to quantify the total amount of exotic matter supporting the wormhole geometry is through the volume integral of the NEC-violating term,
\begin{equation}
I_{\text{ex}} = \int_{r_0}^{\bar r} \mathrm{d} V \, (\rho\left(  r\right)  +p_{r}\left(  r\right))
= - \int_{r_0}^{\bar r} \mathrm{d}r \, 4\pi r^2 \left( \frac{r_{0}\left(  1-B\right)
}{\kappa r^{3}} \right) .
\end{equation}
For all the cases considered in this work, we obtain
\begin{equation}
I_{\text{ex}} = -\frac{4\pi r_0(1-B)}{\kappa}
\ln\left(\frac{\bar r}{r_0}\right).
\end{equation}
This result shows that the exotic matter content depends only logarithmically on the matching radius $\bar{r}$ and is fully determined by the same geometric parameters in all models considered. Moreover, since $\rho+p_r<0$ throughout the integration domain, the NEC is violated in a finite region around the throat. In particular, the total amount of exotic matter decreases as the parameter $B$ approaches unity having in mind that $B \neq 1$, otherwise the flare-out condition is violated. Furthermore, To minimize the effects of the thermal tensor, we have imposed its vanishing at the throat. This condition has allowed us to determine the throat size in terms of the lengths $r_{1}$  and  $r_{2}$. In particular, we have observed that, in each case, the throat is
proportional to $r_{2}$  or to a combination of $r_{1}$ and
$r_{2}$. This behaviour is significant because it allows the size of the throat to be adjusted. Although no external boundary is defined by a vanishing thermal tensor, an external boundary can be identified by examining the profile of the redshift function, at least in the massless case. In particular, when the plates vary, the value of $B_{-}$ must be taken close to unity to place the boundary near the throat. Conversely, when the plates are parametrically fixed, avoiding imaginary solutions for the boundary requires imposing an additional constraint given by
\begin{equation}
d=2\sqrt{r_{1}r_{2}}= \sqrt{\frac{10}{137}}\frac{6N\ell_{\rm P}}{\pi} \,.
\end{equation}
This is not the only possible constraint, but it is the simplest. As we can see, such a condition imposes an additional bound on the value of $r_{2}$, since 
$d$ is fixed by the physical setup. For instance, if the plate
separation $d$ is on the order of  $nm$ $\left(
d\simeq10^{-9}m\right)  $ and $r_{1}$ is Planckian, one gets
$N\simeq10^{26}$.
This result is consistent with that obtained in  \cite{CCW}. For the cases involving potential, a different procedure has been adopted. Since there is no natural boundary arising from the vanishing of the thermal stress tensor, we have investigated the existence of stationary points. Although these points correspond to maximum values of the thermal stress tensor components, the actual values of its components can be very small.
A comment on this point is in order: the significance of identifying an external boundary lies in assessing the validity of the approximation adopted in this work. Indeed, since  we have assumed the Einstein field equations to be valid close to the throat, it is important to understand the range over which this approximation remains valid. Given that our analysis is limited to a narrow region near the throat, we intend to revisit the same field equations in a future paper without relying on this approximation. 

Although the SET in our model contains multiple components,  the present work highlights physically motivated interactions among Casimir energy, thermal stresses, scalar field and electric charge. Key results include the explicit support of the wormhole throat by Casimir plates, the role of thermal anisotropies in maintaining balance, and clear relationships between physically meaningful quantities. These insights provide a framework 
for constructing multi-component wormhole solutions with well-defined physical content.

It should also be emphasized that traversable wormholes supported by Casimir-type effects have been widely studied in the literature under various frameworks. In particular, several works have considered configurations in the absence of additional scalar field contributions, where the exotic matter content is entirely driven by the Casimir energy. In the appropriate limit of vanishing scalar field potential, our results consistently reduce to these previously studied cases.
However, the present construction extends these models by incorporating a physically motivated matter sector composed of three distinct contributions, i.e. electric,  scalar and a thermal contribution. Importantly, no additional or ad hoc SET components are introduced in order to support the wormhole geometry. Instead, the field equations are satisfied entirely within this well-defined physical matter content.
In this way, the model provides a more structured physical interpretation of the source supporting the traversable wormholes, while still allowing for a broad class of solutions. In particular, the total amount of exotic matter can be explicitly quantified through a unified integral expression, which applies uniformly to all cases considered here.

As a final remark, it is worth noting that the resulting wormhole solutions, in both the massless and massive cases, exhibit a throat size that increases with the  strength of the electric field, or equivalently, with the number of elementary charges. It would be interesting though to investigate the full picture by including rotation effects. We leave this topic open for future research.


\end{document}